\documentstyle[12pt,twoside]{article}
 
\def\a{\alpha}
\def\b{\beta}
\def\c{\chi}
\def\d{\delta}
\def\e{\epsilon}                
\def\f{\phi}                    
\def\g{\gamma}
\def\h{\eta}

\def\l{\lambda}
\def\m{\mu}
\def\n{\nu}

\def\p{\pi}                     
\def\r{\rho}                    
\def\s{\sigma}                  
\def\t{\tau}

\def\F{\Phi}
\def\G{\Gamma}

\def\O{\Omega}
\def\P{\Pi}

\def\bo{{\raise.05ex\hbox{\large$\Box$}\:}}             
\def\cbo{{\,\raise-.15ex\Sc [\,}}                       
\def\pa{\partial}                                       
\def\su{\sum}                                           
\def\TH{{\raise.2ex\hbox{$\displaystyle \bigodot$}\mskip-4.7mu \llap H \;}}
\def\face{\hbox{\normalsize$\;\;\:{\raise.9ex\hbox{\oo n}\mskip-13mu \llap
        {${\buildrel{\hbox{\frtnrm ..}}\over\smile}$}}\:$}}     
\def\Face{{\raise.2ex\hbox{$\displaystyle \bigodot$}\mskip-2.2mu \llap {$\ddot
        \smile$}}}                                      
\def\Lhat{{\bf\rlap{\kern-.09em$\hat{\phantom L}$}L}}
\def\Lcheck{{\bf\rlap{\kern-.09em$\check{\phantom L}$}L}}
 
\def\sp#1{{}^{#1}}                              

\def\leftrightarrowfill{$\mathsurround=0pt \mathord\leftarrow \mkern-6mu
        \cleaders\hbox{$\mkern-2mu \mathord- \mkern-2mu$}\hfill
        \mkern-6mu \mathord\rightarrow$}
\def\dvec#1{\vbox{\ialign{##\crcr
        \leftrightarrowfill\crcr\noalign{\kern-1pt\nointerlineskip}
        $\hfil\displaystyle{#1}\hfil$\crcr}}}           
\def\ddt#1{{\buildrel {\hbox{\LARGE .\kern-2pt.}} \over {#1}}}
 
\def\frac#1#2{{\textstyle{#1\over\vphantom2\smash{\raise.20ex
        \hbox{$\scriptstyle{#2}$}}}}}                   
\def\ha{\frac12}                                        
\def\sfrac#1#2{{\vphantom1\smash{\lower.5ex\hbox{\small$#1$}}\over
        \vphantom1\smash{\raise.4ex\hbox{\small$#2$}}}} 
\def\bfrac#1#2{{\vphantom1\smash{\lower.5ex\hbox{$#1$}}\over
        \vphantom1\smash{\raise.3ex\hbox{$#2$}}}}       
\def\afrac#1#2{{\vphantom1\smash{\lower.5ex\hbox{$#1$}}\over#2}}    
 
\def\boxes#1{
        \newcount\num
        \num=1
        \newdimen\downsy
        \downsy=-1.64ex
        \mskip-7.8mu
        \bo
        \loop
        \ifnum\num<#1
        \llap{\raise\num\downsy\hbox{$\bo$}}
        \advance\num by1
        \repeat}
\def\boxup#1#2{\newcount\numup
        \numup=#1
        \advance\numup by-1
        \newdimen\upsy
        \upsy=.82ex
        \mskip7.8mu
        \raise\numup\upsy\hbox{$#2$}}
 
\newskip\humongous \humongous=0pt plus 1000pt minus 1000pt
\def\caja{\mathsurround=0pt}

\newif\ifdtup
\def\panorama{\global\dtuptrue \openup2\jot \caja
        \everycr{\noalign{\ifdtup \global\dtupfalse
        \vskip-\lineskiplimit \vskip\normallineskiplimit
        \else \penalty\interdisplaylinepenalty \fi}}}
\def\li#1{\panorama \tabskip=\humongous                         
        \halign to\displaywidth{\hfil$\displaystyle{##}$
        \tabskip=0pt&$\displaystyle{{}##}$\hfil
        \tabskip=\humongous&\llap{$##$}\tabskip=0pt
        \crcr#1\crcr}}

\def\NP{Nucl. Phys. B\,}
\def\PL{Phys. Lett. B\,}

\def\PRD{Phys. Rev. D\,}
\def\CQG{Class. Quant. Grav.\,}

\def\ref#1{$\sp{#1]}$}

\parskip=\medskipamount                 
\def\baselinestretch{1.2}       
\thicklines                         
\def\title#1#2#3#4{
\begin{document}
        {\hbox to\hsize{#4 \hfill  #3}}\par
        \begin{center}\vskip.5in minus.1in {\Large\bf #1}\\[.5in minus.2in]{#2}
        \vskip1.4in minus1.2in {\bf ABSTRACT}\\[.1in]\end{center}
        \begin{quotation}\par}
\def\author#1#2{#1\\[.1in]{\it #2}\\[.1in]}

\def\AMIC{Aleksandar Mikovic\'c
\\[.1in]{\it Blackett Laboratory, Imperial College, Prince Consort Road, London
SW7 2BZ, UK}\\[.1in]}

\def\AMICIF{Aleksandar Mikovi\'c\,
\footnote{Work supported by MNTRS and Royal Society}
\\[.1in] {\it Blackett Laboratory, Imperial College, Prince Consort
Road, London SW7 2BZ, UK}\\[.1in]
and \\[.1 in]
{\it Institute of Physics, P.O. Box 57, 11001 Belgrade, Yugoslavia}
\footnote{Permanent address}\\ {\it E-mail:\, mikovic@castor.phy.bg.ac.yu}}

\def\AMSISSA{Aleksandar Mikovi\'c\,
\footnote{E-mail address: mikovic@castor.phy.bg.ac.yu}
\\[.1in] {\it SISSA-International School for Advanced Studies\\
Via Beirut 2-4, Trieste 34100, Italy}\\[.1in]
and \\[.1 in]
{\it Institute of Physics, P.O. Box 57, 11001 Belgrade, Yugoslavia}
\footnote{Permanent address}}

\def\AM{Aleksandar Mikovi\'c 
\footnote{E-mail address: mikovic@castor.phy.bg.ac.yu}
\\[.1in] {\it Institute of Physics, P.O.Box 57, Belgrade 11001, Yugoslavia}
\\[.1in]}

\def\AMsazda{Aleksandar Mikovi\'c 
\footnote{E-mail address: mikovic@castor.phy.bg.ac.yu}
and Branislav Sazdovi\'c \footnote{E-mail: sazdovic@castor.phy.bg.ac.yu}
\footnote{Work supported by MNTRS}
\\[.1in] {\it Institute of Physics, P.O.Box 57, Belgrade 11001, Yugoslavia}
\\[.1in]}

\def\AMVR{Aleksandar Mikovi\'c\,
\footnote{E-mail address: mikovic@castor.phy.bg.ac.yu}
\\[.1in] 
{\it Institute of Physics, P.O. Box 57, 11001 Belgrade, Yugoslavia}
\\[.2in]
Voja Radovanovi\'c \\[.1 in]
{\it Faculty of Physics, P.O. Box 550, 11001 Belgrade, Yugoslavia}}

\def\AMCVR{Aleksandar Mikovi\'c
\footnote{Permanent address: Institute of Physics, P.O. Box 57, 11001 
Belgrade, Yugoslavia}\footnote{E-mail: mikovic@fy.chalmers.se, 
mikovic@castor.phy.bg.ac.yu}
\\
{\it Institute of Theoretical Physics, Chalmers University of Technology,
S-412 96 Goteborg, Sweden}\\[.1in]
and
\\[.1in]
Voja Radovanovi\'c
\footnote{E-mail: rvoja@rudjer.ff.bg.ac.yu} \\
{\it Faculty of Physics, P.O. Box 550, 11001 Belgrade, Yugoslavia}}

\def\AMVVR{Aleksandar Mikovi\'c
\footnote{On leave from Institute of Physics, P.O. Box 57, 11001 
Belgrade, Yugoslavia}
\footnote{Supported by Comissi\'on Interministerial de Ciencia y Tecnologia}
\footnote{E-mail: mikovic@lie1.ific.uv.es}
\\
{\it Departamento de Fisica Te\'orica and IFIC, Centro Mixto Universidad
de Valencia-CSIC, Facultad de Fisica, Burjassot-46100, Valencia, Spain}
\\[.1in]
Voja Radovanovi\'c
\footnote{E-mail: rvoja@rudjer.ff.bg.ac.yu} \\
{\it Faculty of Physics, P.O. Box 368, 11001 Belgrade, Yugoslavia}}

\def\AMV{Aleksandar Mikovi\'c
\footnote{On leave from Institute of Physics, P.O. Box 57, 11001 
Belgrade, Yugoslavia}
\footnote{Supported by Comissi\'on Interministerial de Ciencia y Tecnologia}
\footnote{E-mail: mikovic@lie1.ific.uv.es}
\\
{\it Departamento de Fisica Te\'orica and IFIC, Centro Mixto Universidad
de Valencia-CSIC, Facultad de Fisica, Burjassot-46100, Valencia, Spain}}

\def\endtitle{\par\end{quotation}\vskip3.5in minus2.3in\newpage}
 
 
\def\endabstract{\par\end{quotation}
        \renewcommand{\baselinestretch}{1.2}\small\normalsize}
 
 
\def\xpar{\par}                                         

\def\letterhead{
        \centerline{\large\sf INSTITUTE OF PHYSICS}
        \centerline{\sf P.O.Box 57, 11001 Belgrade, Yugoslavia}
        \rightline{\scriptsize\sf Dr Aleksandar Mikovi\'c}
        \vskip-.07in
        \rightline{\scriptsize\sf Tel: 11 615 575}
        \vskip-.07in
        \rightline{\scriptsize\sf E-mail: MIKOVIC@CASTOR.PHY.BG.AC.YU}}

\def\sig#1{{\leftskip=3.75in\parindent=0in\goodbreak\bigskip{Sincerely yours,}
\nobreak\vskip .7in{#1}\par}}

\def\ssig#1{{\leftskip=3.75in\parindent=0in\goodbreak\bigskip{}
\nobreak\vskip .7in{#1}\par}}

 
\def\ree#1#2#3{
        \hfuzz=35pt\hsize=5.5in\textwidth=5.5in
        \begin{document}
        \ttraggedright
        \par
        \noindent Referee report on Manuscript \##1\\
        Title: #2\\
        Authors: #3}
 
 
\def\start#1{\pagestyle{myheadings}\begin{document}\thispagestyle{myheadings}
        \setcounter{page}{#1}}
 
 
\catcode`@=11
 
\def\ps@myheadings{\def\@oddhead{\hbox{}\footnotesize\bf\rightmark \hfil
        \thepage}\def\@oddfoot{}\def\@evenhead{\footnotesize\bf
        \thepage\hfil\leftmark\hbox{}}\def\@evenfoot{}
        \def\sectionmark##1{}\def\subsectionmark##1{}
        \topmargin=-.35in\headheight=.17in\headsep=.35in}
\def\ps@acidheadings{\def\@oddhead{\hbox{}\rightmark\hbox{}}
        \def\@oddfoot{\rm\hfil\thepage\hfil}
        \def\@evenhead{\hbox{}\leftmark\hbox{}}\let\@evenfoot\@oddfoot
        \def\sectionmark##1{}\def\subsectionmark##1{}
        \topmargin=-.35in\headheight=.17in\headsep=.35in}
 
\catcode`@=12
 
\def\sect#1{\bigskip\medskip\goodbreak\noindent{\large\bf{#1}}\par\nobreak
        \medskip\markright{#1}}
\def\chsc#1#2{\phantom m\vskip.5in\noindent{\LARGE\bf{#1}}\par\vskip.75in
        \noindent{\large\bf{#2}}\par\medskip\markboth{#1}{#2}}
\def\Chsc#1#2#3#4{\phantom m\vskip.5in\noindent\halign{\LARGE\bf##&
        \LARGE\bf##\hfil\cr{#1}&{#2}\cr\noalign{\vskip8pt}&{#3}\cr}\par\vskip
        .75in\noindent{\large\bf{#4}}\par\medskip\markboth{{#1}{#2}{#3}}{#4}}
\def\chap#1{\phantom m\vskip.5in\noindent{\LARGE\bf{#1}}\par\vskip.75in
        \markboth{#1}{#1}}
\def\refs{\bigskip\medskip\goodbreak\noindent{\large\bf{REFERENCES}}\par
        \nobreak\bigskip\markboth{REFERENCES}{REFERENCES}
        \frenchspacing \parskip=0pt \renewcommand{\baselinestretch}{1}\small}
\def\unrefs{\normalsize \nonfrenchspacing \parskip=medskipamount}
\def\Item{\par\hang\textindent}
\def\Itemitem{\par\indent \hangindent2\parindent \textindent}
\def\makelabel#1{\hfil #1}
\def\topic{\par\noindent \hangafter1 \hangindent20pt}
\def\Topic{\par\noindent \hangafter1 \hangindent60pt}

\title{One-loop Effective Action for Spherical Scalar Field Collapse}
{\AMVVR}{FTUV/97-35, IFIC/97-34}{June 1997}

We calculate the complete one-loop effective action for a spherical scalar
field collapse in the large radius approximation. This action gives the 
complete trace anomaly, which beside the matter loop contributions, receives 
a contribution from the graviton loops. Our result opens a possibility for a 
systematic study of the back-reaction effects for a real black hole.

\endtitle

\sect{1. Introduction}

In reference \cite{mr4} a background-field formalism has been set up for
calculating the complete one-loop effective action for a generic 2d dilaton 
gravity whose potential has a certain asymptotic behavior. This asymptotics was
taken because it appears in the 2d dilaton gravity models which describe
the spherical general relativity \cite{s,cmn}, as well as in the CGHS model of 
2d black holes \cite{cghs}. Therefore the effective action derived in
\cite{mr4} can describe the back-reaction effects for a realistic 4d black 
hole.
However, the matter in \cite{mr4} does not couple to the dilaton, so that
the action derived there corresponds to a spherical null-dust collapse, which 
is not the most realistic model of collapse, although it is useful, since the
classical equations of motion can be integrated \cite{msnd}, and 
consequently one can develop an operator quantization by using the techniques
developed for the CGHS case \cite{m95,m96,mr}.

In this paper we apply the formalism of \cite{mr4} to the case when the
dilaton is coupled to the matter, in order to obtain a one-loop effective
action for a spherical scalar field collapse. In ref. \cite{eno} a similar 
method has been applied to the case of general matter-dilaton coupling, but 
only the divergent part of the effective action has been calculated.

We start from the Einstein-Hilbert action in 4d with a minimally coupled
scalar field $f$
$$S=\int d^4x \sqrt{-g_4} \left({1\over 16\p G} R_4 - \ha (\nabla_4 f)^2 \right)
\quad.\eqno(1.1)$$
By using a spherically symmetric reduction ansatz \cite{s}
$$ ds^2=\tilde g_{\m\n}dx^{\m}dx^{\n}+ e^{-2\F}d\O ^2 \quad, \eqno(1.2)$$
one obtains
$$ S =  \int d^2 x \sqrt{-{\tilde g}} e^{-2\F}\left( \tilde R
 + 2 ( \tilde \nabla \F)^2 +2e^{2\F} - \ha (\nabla f)^2 \right) \quad ,
\eqno(1.3)$$
where we have set the Newton constant $G$ to one. For the purpose of obtaining
the semi-classical limit, it is useful to replace the scalar field with
$N$ scalar fields, so we will consider the action
$$ S =  \int d^2 x \sqrt{-{\tilde g}}\left[ e^{-2\F}\left( \tilde R
 + 2 ( \tilde \nabla \F)^2 + 2e^{2\F} \right) - {1\over 2}\su_{i}
e^{-2\F}( \tilde \nabla f_i )^2 \right]\quad .\eqno(1.4)$$
The calculation of the effective action simplifies if one performs
a conformal transformation ${\tilde g}_{\m\n}=e\sp{\F}g_{\m\n}$, which gives
$$ S =  \int d^2 x \sqrt{ -g }\left[ \f R
 + 2\f ^{-{1\over 2}} -
  {1\over 2}\f \sum_{i}( \nabla f_i )^2 \right]\quad ,
\eqno(1.5)$$
where $\f =e^{-2\F}$. The action (1.5) will be our classical action.

\sect{2. Background field method}

The one loop effective action for the classical action (1.5) can be found by
using the background field method developed in \cite{mr4}. In \cite{mr4}
a one-loop effective action for a spherical null-dust collapse has been found, 
in the limit of large radius $r = \sqrt\f = e\sp{-\F} \gg 1$. Since the 
spherical null-dust action differs from (1.5) only in the $\f$-dependent matter
coupling, the corresponding calculation for (1.5) is going to be almost 
identical, except for appropriate modifications due to the $\f$-dependent
matter coupling in (1.5).

The one-loop effective action will be given by 
$$\G_1 [\f_0] =S(\f _0)-{1 \over 2i} {\rm Tr} 
\left(\log S^{\prime \prime }(\f _0)\right)\quad ,
\eqno(2.1)$$
where $S^{\prime \prime}(\f _0)$ is the second functional derivative of the 
classical action evaluated for the set of classical background fields
$\f_0 = \{g_{\m\n},\  \f,\ f_0\}$. The corresponding quantum fields are 
denoted as $\{h_{\m\n},\ \hat \f,\  f\}$.
We choose the same gauge-fixing condition as in the null-dust case  
$$\c _{\m}=D_{\l}h^{\l}_{\m}-{1\over 2}D_{\m}h -{1\over \f }D_{\m}\hat \f =0
\quad,\eqno(2.2)$$
which produces the same gauge-fixing term in the action 
$$S_{GF}=-{1\over 2}\int d\sp 2 x\sqrt{-g}\f \c _{\m}\c ^{\m}\quad.\eqno(2.3)$$
The effect of (2.3) is that the new action has a minimal structure, i.e. 
the second spacetime derivatives acting on the quantum fields appear only 
as $\Box$. We also rescale the quantum fields as
$$ h_{\m\n} \to {h_{\m\n}\over\sqrt\f} \quad,\quad {\hat \f} \to 
{\sqrt\f} \hat \f
\quad,\quad f \to {f\over\sqrt\f} \quad,\eqno(2.4)$$
in order to remove the $\f$ dependence from the
kinetic terms for the quantum fields. The Jacobian of the transformation
(2.4) is equal to 1. After that we use the t'Hooft-Veltman 
complexification of fields \cite{thv} and take the spacetime dimension to be
$D = 2 + \e$, in order to be able to use the dimensional regularization 
procedure. The quadratic part of the action is then given by
 $$S^{(2)}_{tot}={1\over 2}\int d\sp 2 x \sqrt{-g} 
(\bar h^{*\m\n}\ h^*\ \hat \f ^* \ f^*)
\hat K(I\Box +\hat K^{-1}\hat M)
 \left( \matrix {\bar h_{\r\s}\cr
h\cr \hat \f \cr f\cr }\right)\quad, \eqno(2.5)$$
where $I=diag (\P^{\m\n}_{\r\s},\ 1,\ 1,\ 1)$, 
${\bar h}_{\m\n} = \P_{\m\n}\sp{\r\s} h_{\r\s} = h_{\m\n} - {1\over D} h g_{\m\n}$
and 
$$\hat K=  \left( 
\matrix {\P^{\r\s}_{\m\n}& 0&0&0\cr
0&-{\e \over 2(2+\e )} &-1&0\cr
0&-1 &2 &0\cr
0&0&0&2 \cr }\right),\ \eqno(2.6)
$$    
$$ \hat K^{-1}\hat M= \left( \matrix {\hat V^{\r\s}_{\a\b}&
\hat G_{\a\b}&\hat H_{\a\b}&\hat W_{\a\b}\cr
\hat M^{\r\s}&\hat P&\hat Q&\hat X\cr
\hat N^{\r\s}&\hat L&\hat S&\hat E \cr
\hat Y^{\r\s}&\hat Z&\hat O&\hat F \cr }\right)\quad.\eqno(2.7)$$  
The  matrix elements in (2.7) are given by
$$\li {\hat V_{\a\b}^{\r\s}=&-\P _{\a\b}^{\r\s}R+\P_{\a\b}^{\m\n}(R^{\r}_{\m}
\d ^{\s}_{\n}+R^{\s}_{\m}\d ^{\r}_{\n}-R^{\r\ \ \s}_{\ \m\n \  }
-R^{\r\ \ \s}_{\ \n\m\  })\cr
&-3\P^{\r\s}_{\a\b}\Box \F+7\P^{\r\s}_{\a\b}(\nabla
\F)^2+
2D_{\l} \F(\P^{\s a}_{\a\b}g^{\l\r}+\P^{\r a}_{\a\b}g^{\l\s}\cr
&-\P^{\l\s}
_{\a\b}g^{a\s}-\P^{\l\r}_{\a\b}g^{a\s}){\overrightarrow D}_a
+4\P ^{\m\n}_{\a\b}
(\d ^{\r}_{\m}D^{\s}D_{\n}\F\cr
 &-2\d ^{\r}_{\m}D^{\s} \F D_{\n}\F +\d
^{\s}_{\m} D^{\r}D_{\n}\F  -2\d ^{\s}_{\m}D^{\r}\F D_{\n}\F)\cr
&-\P^{\m\n}_{\a\b}(\d ^{\r}_{\m}D _{\n}f_0D ^{\s}f_0+
\d ^{\s}_{\m}D _{\n}f_0D ^{\r}f_0)+{1\over 2}\P ^{\r\s}_{\a\b}(\nabla f_0)^2-
2\P_{\a\b}^{\r\s}\f^{-{3\over 2}}
,
&(2.8)\cr }$$
$$\li {\hat G_{\a\b}=&\P^{\m\n}_{\a\b}\left({2-\e \over 2+\e } 
R_{\m\n}+2{3\e -2\over
2+\e }D_{\m}\F D_{\n}\F 
-2{\e  -2\over 2+\e }D_{\m}D_{\n}\F +2D_{\m} \F
 {\overrightarrow D}_{\n}\right)\cr
 &+{\e -2\over 2(2+\e)}\P^{\m\n}_{\a\b}D_{\m}f_0D_{\n}f_0
  , &(2.9)\cr }$$    
$$\hat H_{\a\b}=\P ^{\m\n}_{\a\b}\left( D_{\m}f_0D_{\n}f_0-2R_{\m\n}\right)
,\eqno(2.10)$$
$$\hat W_{\a\b}=2\P ^{\m\n}_{\a\b}(1+e^{\F })\pa _{\n}f_0
{\overrightarrow \pa}_{\m}+2\P ^{\m\n}_{\a\b}D_{\m}f_0D_{\n}\F \quad, 
\eqno(2.11)$$
$$\li {\hat M^{\r\s}&={2\e \over 1+\e }R^{\r\s}-{2+\e \over 1+\e }
\left(2{3\e -2\over 2+\e }D^{\r}\F D^{\s}\F 
-2 {\e -2\over \e +2}D^{\r}D^{\s}\F +(
\overleftarrow{\pa^\r}D^{\s}\F  +\overleftarrow{\pa^\s} D^{\r}\F )\right)\cr
&-{\e \over 1+\e }D^{\r}f_0D^{\s}f_0,
&(2.12)\cr }$$
$$ \li {\hat P=&-{\e ^2\over (1+\e )(2+\e )}R -
{2+\e \over 1+\e }{\Big(}-D^{\m} \F \vec
\pa _{\m}-{-\e ^2 +5\e +6\over (2+\e )^2 }\Box\F\cr
&+{-4\e ^2 +3\e +6\over
(2+\e )^2}(\nabla \F )^2+{7\e-2\over 2(2+\e)}\f^{-{3\over2}}
{\Big)} +{\e ^2\over 2(1+\e)(2+\e)}(\nabla f_0)^2, &(2.13)\cr }$$
$$\hat Q={\e \over 2(1+\e ) }((\nabla f_0)^2-2R)
+{2+\e \over 1+\e }(2 \overleftarrow{\pa^\m} D_{\m}
\F +2\Box
 \F +2(\nabla \F )^2-{1\over2}\f^{-{3\over2}}), \eqno(2.14)$$
 $$\hat X= (2-e^{\F }{\e \over 1+\e })\pa_{\m}f_0 {\overrightarrow \pa}^{\m}
+2D^{\n}\F D_{\n}f_0 ,
 \eqno(2.15)$$
$$\li{\hat N^{\r\s}=&-{1 \over 1+\e }R^{\r\s}-{2+\e \over 2(1+\e )}{\Big(} 2
 {3\e -2\over 2+\e }D^{\r}\F D^{\s}\F 
-2 {\e -2\over \e +2}D^{\r}D^{\s}\F \cr
 &+(\overleftarrow{\pa^\r}D^{\s}\F  + \overleftarrow{\pa^\s} D^{\r}\F ){\Big)}
 +{1\over 2(1+\e)}D^{\r}f_0D^{\s}f_0
\quad,&(2.16)\cr}$$
$$\li {\hat L&={\e \over2(2+\e  )(1+\e )}R  -{\e \over 2(1+\e )}D^{\m}\F 
{\overrightarrow \pa}_{\m}
+{2-\e \over 2+\e }\Box \F\cr
 &-{-3\e ^2+6\e +8\over 2(1+\e )(2+\e )}(\nabla \F )^2-{\e \over 4(1+\e)(2+\e)}
 (\nabla f_0)^2-{9\e\over 4(1+\e)}\f^{-{3\over2}} , &(2.17)\cr }$$
$$\hat S= -{\e \over 2(1+\e )}R+
{2+\e \over 1+\e }(\overleftarrow{\pa^\m}D_{\m}\F+{1\over2}\f^{-{3\over2}})
+{1\over
1+\e }\left( \Box \F  +(\nabla \F )^2 
+{\e \over 4}(\nabla f_0)^2+{3\e\over 4}\f^{-{3\over2}}\right), \eqno(2.18)$$
$$\hat E={\e \over 2(1+\e )}e^{\F }\pa ^{\m}f_0\vec \pa _{\m},
\eqno(2.19)$$
$$\hat Y^{\r\s}=\overleftarrow{\pa^\r}(e^{\F
}+1)\pa^{\s}f_0+D^{\r}f_0D^{\s}\F ,\eqno(2.20)$$
$$ \hat Z=-{\e  \over 2(2+\e )}\overleftarrow{\pa^\r} (1+e^{\F })\pa _{\r} f_0
-{\e \over 2(2+\e)}D^{\r}\F D_{\r}f_0 \quad,
\eqno(2.21)$$
$$\hat O=-\overleftarrow{\pa^\r}\pa _{\r}f_0-D^{\r}f_0D_{\r}\F \quad, \eqno(2.22)$$
$$\hat F=\Box \F-(\nabla \F)^2\quad. \eqno(2.23)$$
The novel features in the spherical scalar case are that $\hat O$ and $\hat F$
matrix elements are non-zero, while the other matrix elements are modified by
the terms coming from the dilaton-matter coupling. 

After fixing of the quantum gauge symerties, we must add the ghost
action to the action (2.5). The ghost action is the same as in the
null-dust case \cite{mr4}
$$S_{gh}=\int d^2x\ \sqrt{-g} \f\bar c^{\m}\ [-\Box c_{\m}-R^{\n
}_{\m}c_{\n}+{1\over \f }D_{\m}(c^{\r}\pa _{\r}\f )]\quad. \eqno(2.24)$$
In (2.24) we have omitted the terms which do not contribute to the one-loop 
effective action. We then rescale the ghosts as
$$\f \bar c^{\m}\rightarrow \bar c^{\m}\quad,\eqno(2.25)$$
so that
$$S_{gh}=\int d^2x \sqrt{-g} \bar c^{\m} 
[-\Box c_{\m}-R^{\n}_{\m}c_{\n}+{1\over \f }D_{\m}(c^{\r}\pa _{\r}\f )] 
\quad. \eqno(2.26)$$

\sect{3. Expansion around a flat metric }

The calculation of the one-loop effective action can be simplified  
by expanding the
background metric around a flat metric $\h_{\m\n}$ as
$$g_{\m\n} = \h_{\m\n} +\g_{\m\n} + O(\g^2 ) \quad. \eqno(3.1)$$
After inserting (3.1) into (2.6) and (2.7), we get
$$\sqrt{-g} (I\Box +\hat K^{-1}\hat M)=diag (P^{\r\s}_{\a\b},\, 1,\, 1,\, 1)\pa^2
 +  K^{-1} M \, ,\eqno(3.2)$$ 
where $\pa ^2 =\h ^{ab}\pa _a\pa _b$, $P^{\r\s}_{\a\b}$ is given in (3.8)
and
$$K^{-1} M=\left(\matrix{{\tilde V}^{\r\s}_{\a\b}&G_{\a\b}&H_{\a\b}&W_{\a\b}\cr
M^{\r\s}&P&Q&X\cr
N^{\r\s}&L&S&E \cr
Y^{\r\s}&Z&O&F \cr }\right)\quad .\eqno(3.3)$$
The matrix elements in (3.3) which are relevant for our calculation
are given by
$$\li {{\tilde V}_{\a\b}^{\r\s}=&\overleftarrow{\pa_a}A^{ ab\r\s}_{\a\b}{\overrightarrow \pa}_{b}\cr
-&\left(\h ^{ab}
P^{\r\s}_{\e \t }S^{\e \t}_{b\a\b}
-2\sqrt{-g} \pa _{\l}\F (g^{\l\r}\P ^{\s a}_{\a\b}+
\P ^{\r a}_{\a\b}g^{\l\s}-\P^{\l\s}_{\a\b}g^{a\r}-\P ^{\l\r}_{\a\b}g^{a\s} )
\right) {\overrightarrow \pa}_a  \cr
+&\overleftarrow{\pa_a} \h^{ab}P^{\e\t}_{\a\b}S^{\r\s}_{b\t\e }-\sqrt{-g}
 \P _{\a\b}^{\r\s}R+\sqrt{-g} \P _{\a\b}^{\m\n}(R^{\r}_{\m}
\d ^{\s}_{\n}+R^{\s}_{\m}\d ^{\r}_{\n}-R^{\r\ \ \s}_{\ \m\n \  }
-R^{\r\ \ \s}_{\ \n\m\  })\cr
-&3\sqrt{-g} \P^{\r\s}_{\a\b}\Box \F +7\sqrt{-g} \P^{\r\s}_{\a\b}(\nabla
 \F )^2+\cr 
+&4\sqrt{-g} \P ^{\m\n}_{\a\b}
\left( \d ^{\r}_{\m}D^{\s}D_{\n}\F -2\d ^{\r}_{\m}D^{\s} \F D_{\n}\F +\d
 ^{\s}_{\m} D^{\r}D_{\n}\F  -2\d ^{\s}_{\m}D^{\r}\F D_{\n}\F \right)  \cr
-&2\sqrt{-g} \pa_{\l}\F (\P^{a\t}_{\a\b}g^{\l\e}+\P^{\e a}_{\a\b}g^{\l
\t}\cr 
-&\P^{\l\e}_{\a\b}g^{a\t}-\P^{\l\t}_{\a\b}g^{a\e})S^{\r\s}_{a\e\t} +\h
^{ab}P^{\e\t}_{\d\h}S^{\d\h}_{a\a\b}S^{\r\s}_{b\e\t}\cr
-&\sqrt{-g}\P^{\m\n}_{\a\b}(\d ^{\r}_{\m}D _{\n}f_0D ^{\s}f_0+
\d ^{\s}_{\m}D _{\n}f_0D ^{\r}f_0)+{1\over 2}\sqrt{-g}\P ^{\r\s}_{\a\b}(
(\nabla f_0)^2  -4\f ^{-{3\over2}}) 
\,,&(3.4)\cr }$$
$$\li{P=&\overleftarrow{\pa_a}\bar \g^{ab}{\overrightarrow \pa}_b 
-{\e ^2\over (1+\e )(2+\e )}\sqrt{-g} R 
-{2+\e \over 1+\e } \sqrt{-g}{\Big(} -D^{\m} \F  
{\overrightarrow \pa}_{\m}\cr 
 &- {-\e ^2 +5\e +6\over (2+\e )^2 }\Box \F +{-4\e ^2 +3\e +6\over
(2+\e )^2}(\nabla \F )^2+{7\e-2\over 2(2+\e )}
\f^{-{3\over2}}{\Big)}\cr
&+{\e ^2\over 2(1+\e)(2+\e)}\sqrt{-g}(\nabla f_0)^2\quad,
&(3.5)\cr }$$
$$\li{S=& {\overleftarrow \pa}_a\bar \g^{ab} {\overrightarrow \pa}_b 
-{\e \over 2(1+\e )}\sqrt{-g} (R-{1\over 2}(\nabla f_0)^2)+{4+5\e\over 4(1+\e)}
\f^{-{3\over2}}\cr 
&+\sqrt{-g} {1\over 1+\e }\left( (2+\e ) \overleftarrow{\pa^\m} D_{\m}\F 
+\Box \F  +(\nabla \F )^2\right) \quad, & (3.6)\cr }$$
$$F={\overleftarrow \pa}_a \bar \g^{ab} {\overrightarrow \pa}_b +
\sqrt{-g} (\Box \F -(\nabla \F)^2)\quad,
\eqno(3.7)$$
where
$$ P^{\m\n}_{\r\s}={1\over 2} (\d^{\m}_{\r}\d^{\n}_{\s}+\d^{\m}_{\s}
\d^{\n}_{\r}
)-{1\over D}\h^{\m\n}\h_{\r\s},$$
$$S^{\r\s}_{a\m\n}=2\G^{(\r}_{a(\m}\d^{\s)}_{\n)},$$
$$\bar \g^{\m\n} = \g^{\m\n}-{1\over 2}\g \h^{\m\n},$$
$$A^{\r\s ab}_{\a\b}=P^{\r\s}_{\a\b}\bar \g^{ab}-{1\over D}\h^{ab}
(\g^{\r\s}\h_{\a\b}-\g_{\a\b}\h^{\r\s})
\quad, \eqno(3.8)$$
and $\g =\g^{\m\n}\h_{\m\n}$. 
In the case of the ghost action (2.26), the expansion (3.1) yields \cite{mr4}
$$\li{S_{gh}=&\int d^2x\ \bar c^{\m} (\d^{\n}_{\m}\pa^2 +T^{\n}_{\m})c_{\n}\cr
=&\int d\sp 2 x\bar c^{\m}{\Big[} \d^{\n}_{\m}\pa^2+\d^{\n}_{\m}{\overleftarrow 
\pa}_a
{\bar \g}^{ab}
{\overrightarrow \pa}_{b}-\G^{\n}_{a\m} \h^{a\s} {\overrightarrow\pa}_{\s} +
\h^{ab}\G^{\r}_{a\m}
\G^{\n}_{b\r}\cr 
&+{\overleftarrow\pa}_{\s}\h ^{a\s}\G
^{\n}_{a\m}
+2\pa _{\r}\F (\h ^{\r\n}-{\bar \g}^{\r\n}){\overrightarrow \pa}_{\m}
 -2\h ^{\r\a}\G ^{\n}_{\a\m}\pa _{\r}\F +\sqrt{-g} R_{\m}^{\n}\cr
 &-2\d ^{\r\n}\G ^{\a}_{\r\m}\pa _{\a} \F - (\h ^{\r\n}-\bar \g
^{\r\n})( 4\pa _{\m}\F \pa _{\r}\F  -2\pa _{\m}\pa _{\r}\F ){\Big]} c_{\n}
\quad.&(3.9)\cr }$$
The one-loop correction to the effective action is then given by
$$\G _1={i\over 2}{\rm Tr} \log \left( 1+K^{-1}M{1\over  \pa^2}\right)
-i{\rm Tr} \log \left( 1+T{1\over \pa^2}\right), \eqno(3.10)$$
where $K^{-1}M$ and $T$ are defined by (3.3) and (3.9).   
After expanding the logarithm in (3.10), we obtain
$$\li{\G_1 =&{i\over 2}tr{\Big [}({\tilde V}^{\r\s}_{\a\b}P^{\a\b}_{\r\s}+P+S+F)
{1\over \pa^2}-{1\over 2}
{\Big(}{\tilde V}^{\r\s}_{\a\b}{1\over \pa^2}P^{\a\b}_{\m\n}{\tilde V}^{\m\n}_{\g\d}
{1\over \pa^2}P^{\g\d}_{\r\s}\cr
&+2G_{\a\b}{1\over \pa^2}P^{\a\b}_{\m\n}M^{\m\n}{1\over \pa^2}
+2H_{\a\b}{1\over \pa^2}P^{\a\b}_{\m\n}N^{\m\n}{1\over \pa^2}+\cr
&+2Y_{\a\b}{1\over \pa^2}P^{\a\b}_{\m\n}W^{\m\n}{1\over \pa^2}+
P{1\over \pa ^2}P{1\over \pa ^2}+
S{1\over \pa ^2}S{1\over \pa ^2}\cr
 &+F{1\over \pa ^2}F{1\over \pa ^2}
+2Q{1\over \pa ^2}L{1\over \pa ^2}+
2X{1\over \pa ^2}Z{1\over \pa ^2} + 2E{1\over \pa ^2}O{1\over \pa ^2}
{\Big)}{\Big]}\cr
&-i\, tr{\Big[}T^{\m}_{\n}\d ^{\n}_{\m}{1\over  \pa ^2}-{1\over 2}
T^{\m}_{\n}\d ^{\n}_{\r}
{1\over  \pa ^2} 
T^{\r}_{\s}\d ^{\s}_{\m}{1\over  \pa^2}{\Big]}\quad,\ &(3.11)\cr }$$
where $tr$ denotes the spacetime trace, and symmetric ordering has been taken
in the vertices, i.e. $v(x)p \to \ha (v(x)p + p v(x))$ where $p=i\pa_x$.

\sect{4. One-loop diagrams and the effective action}
 
As in the null-dust case \cite{mr4}, we will calculate (3.11) by evaluating it 
for $g_{\m\n}=\h_{\m\n}$ and for $\F$=constant and then add these 
contributions to the contribution which vanishes in these special cases. 

The contribution due to $\bar h_{\m\n}$ in the loops can
be written as
$$\li{tr\left( {\tilde V}{1\over \pa ^2}-{1\over 2}
{\tilde V}{1\over \pa ^2}{\tilde V}{1\over \pa ^2}\right)
=& tr\left[ (A+B+C){1\over \pa ^2} \right]\cr
&-{1\over 2} tr\left( A{1\over \pa ^2}A{1\over \pa ^2}+
B{1\over \pa ^2}B{1\over \pa ^2}+C{1\over \pa ^2}C{1\over \pa ^2} \right)\cr
 &-tr\left( A{1\over \pa ^2}B{1\over \pa ^2}+ A{1\over \pa ^2}C{1\over \pa ^2}
+ B{1\over \pa ^2}C{1\over \pa ^2}\right) \quad.&(4.1)\cr }$$
In (4.1), we denote the vertices with two, one and zero spactetime
derivatives as A, B, and C, respectively. We will also refer to terms 
$tr\,(X{1\over \pa ^2})$ and $tr\,(X{1\over \pa ^2}Y{1\over \pa ^2})$ as 
diagrams $X$ and $XY$ respectively, where $X$ and $Y$ are any of the vertices.

It is easy to see that $A=B=0$ after the infrared regularization (see
the appendix of \cite{mr4}). The $C$ diagram is given by
$$\li {C =& -{i\p ^{{D\over 2}}\over (2\p )^2}\G (-{\e \over 2})
\int d\sp 2 x\ \left[ \left({-D^2+D+2\over 2}
 -{4\over D}\right)\sqrt{-g} R+\h ^{ab}P^{\e\t}_{\d\h}P^{\a\b}_{\r\s}
 S^{\d\h}_{a\a\b}
S^{\r\s}_{b\e \t}\right]  \cr 
&+4{i\p^{{D\over 2}}\over (2\p)^2}\G (-{\e \over 2}){D+2\over 2}\int
d\sp 2 x \sqrt{-g} R\F \cr
&-{i\p^{{D\over  2}}\over (2\p )^2}
(D^2+D-2)\G(-\e/2) \left(({7\over 2}
-{8\over D})\int d\sp 2 x \sqrt{-g}{\Big[}(\nabla \F )^2 -
{N+2\over 2}\Box \F {\Big]}\right. \cr
&+\left. ({1\over 4}-{1\over
D})N\int d\sp 2 x\sqrt{-g}(\nabla f_0)^2-\int d^2x\sqrt g\f^{-{3\over2}}\right)
.&(4.2)\cr }$$
The term with $f_0$ in (4.2) is new, and appears due to the dilaton-matter 
coupling. The factor $N$ comes from the $N$ scalar
fields. As explained in \cite{mr4}, the $AA$ diagram is given by 
$$- {i\pi ^{{D\over 2}}\over (2\p )^2} {D^2+D-2\over 2}\G (1-{\e \over 2})
B(2+{\e \over 2}, 2+{\e \over 2}) \int d\sp 2 x\sqrt{-g}\left(R{1\over \Box }R+
{4\over \e(1+{\e \over 2})} R\right), \eqno(4.3)$$
while the $AB$ diagram vanishes. The $AC$ diagram is given by
$$AC=2{i\p ^{{D\over 2}}\over (2\p )^2} \int d\sp 2 x\sqrt{-g}\left( 
R{1\over \Box }R+{N\over 2}R{1\over \Box}(\nabla f_0)^2+R{1\over \Box}(\nabla \F)^2
-R\F \right). \eqno(4.4)$$ 
For the $BB$ diagram we get
$$\li {BB=&-8{i\p ^{{D\over 2}} \over (2\p )^2}\left( \int d\sp 2 x \sqrt{-g}
R{1\over \Box }R+4\int d\sp 2 x\sqrt{-g}R\right) \cr
&-2{i\p ^{{D\over 2}}\over (2\p )^2 }\G (-{\e \over 2})\int d\sp 2 x  
[\h ^{ab}P^{\e\t}_{\d \h}P^{\a\b}_{\r\s}S^{\d\h}_{a\a\b}
S^{\r\s}_{b\e \t}-2(1-\e )(D+2)\sqrt{-g} R\F ]\cr
&+4{i\p^{{D\over 2}}\over(2\p )^2}\G (-{\e \over 2})B(1+{\e
\over 2},1+{\e \over 2})(D^2+D-2)\int d\sp 2 x \sqrt{-g}(\nabla \F )^2 . 
&(4.5)\cr }$$
The non-covariant terms in (4.2) and (4.5) vanish. The $BC$ and $CC$
diagrams are infrared divergent, but after an appropriate regularization 
\cite{mr4}, they also vanish.

The contribution to the effective action from the $P$ and $PP$ diagrams are
the same as in the null-dust case.
This is a consequence of the fact that $P = O(\e ^2 )$. On the
other hand, the $S$ and the $F$ diagram have a non-zero contribution to the 
effective action. If we denote as $X$ a diagram in the set $ \{ P,\ S,\ F\}$, 
then
$$\li{\sum_X tr{\Big(} X {1\over \pa ^2}-{1\over 2}X{1\over \pa ^2}&X{1\over
\pa ^2}{\Big)} = -{i\p ^{{D\over 2}}\over  (2\p )^2} 
\G(-{\e \over 2})
 {4\e ^2-3\e-6\over  (1+\e  )(2+\e )}\int d\sp 2 x
 \sqrt{-g}(\nabla  \F)^2\cr
&-{i\p ^{{D\over 2}}\over  (2\p )^2}\G (-{\e \over2}) 
\left({1\over 1+\e }
-{1\over 2}\left( {2+\e \over 1+\e }\right)^2\right)
\int d\sp 2 x\sqrt{-g}(\nabla \F )^2\cr
&-4{i\p^{{D\over 2}}\over(2\p )^2}\G (-{\e \over2})
\int d\sp 2 x\sqrt{-g}\Box \F \cr 
&+ {N+2\over 2}{i\p^{{D\over 2}}\over (2\p )^2}\G (1-{\e \over
2})B(2+{\e \over 2},2+{\e \over 2})
{\Big(} \int d\sp 2 x\sqrt{-g}  R{1\over \Box }R \cr 
&+{4\over \e(1+{\e \over 2})}\int d\sp 2 x\sqrt{-g}R {\Big)}
 +{i\p^{{D\over 2}}\over (2\p )^2}
\int d\sp 2 x \sqrt{-g} R\F \cr
&-2 {i\p ^{{D\over 2}}\over (2\p )^2}\int d\sp 2 x\sqrt{-g} R{1\over
\Box}(\nabla \F )^2 +N{i\p ^{D\over 2}\over (2\p )^2}\G (-{\e \over 2})
\int d\sp 2 x \sqrt{-g}
(\nabla \F)^2
\cr
 &+{N\over 2}{i\p ^{D\over 2}\over (2\p )^2}\int d\sp 2 x \sqrt{-g} 
 \left(-2R{1\over \Box}(\nabla \F )^2+2R\F +(\nabla f_0)^2 \right)\cr
 &+{i\p^{{D\over2}}\over (2\p)^2}\G(-{\e\over2}){9\e-8\over4(1+\e)}\int
d^2x\sqrt g\f^{-{3\over2}}
  .&(4.6)\cr}$$

The diagrams $XZ$ and $EO$ are combinations of $(\nabla f_0)^2$
and $e^{\F}(\nabla f_0)^2$, and they can be neglected  in the large-radius
limit. The diagram $WY$ is divergent, and it is given by
$$ {D\sp 2 + D -2\over 2D} N
{i\p^{D \over 2}\over 4\p\sp 2 } \G (-{\e \over 2})
B(1+{\e \over 2},1+{\e \over 2})\int d\sp 2 x\sqrt{-g}
 (\nabla f_0 )^2 \quad.$$
The diagrams $GM$ and $QL$ are the same as in the null-dust case, and they
are given by
$$\li{GM=-4{i\p^{{D\over 2}}\over (2\p )^2}&{2+\e \over 1+\e }
{\Big(}{1\over 2}
\G (-{\e  \over 2})B(1+{\e \over 2},1+{\e \over 2}){D^2+D-2\over 2D}\cr
&-\G (1-\e/ 2)(1-1/D){\Big)}
\cdot\int d\sp 2 x \sqrt{-g}(\nabla\F )^2 \quad, \cr 
QL=2{i\p ^{{D\over  2}}\over (2\p )^2}&\int d\sp 2 x\sqrt{-g}
(\nabla \F)^2 \quad.&(4.7)\cr}$$

The contribution to the effective action due to the ghost loops 
is given by
$$\li { tr\left( T{1\over  \pa^2}-{1\over 2}  T
{1\over  \pa^2} T{1\over  \pa^2}\right) =&
 tr\left[(\bar A+\bar B+\bar C){1\over \pa ^2}\right]\cr
-&{1\over 2}tr\left(\bar A{1\over \pa ^2}\bar A{1\over \pa ^2}+
\bar B{1\over \pa ^2}\bar B{1\over \pa ^2}+\bar C{1\over \pa ^2}\bar C
{1\over \pa^2} \right)\cr
-& tr\left(\bar A{1\over \pa ^2}\bar B{1\over \pa ^2}+\bar A{1\over \pa ^2}\bar C
{1\over \pa^2}+ \bar B{1\over \pa ^2}\bar C{1\over \pa ^2}\right)
\quad.&(4.8)\cr }$$
The diagrams which appear in (4.8) are the same as in the null-dust case,
so that
$$\bar A=\bar B=\bar A\bar B=\bar B\bar C=0,\eqno(4.9) $$
$$\li {\bar C&
=-{i\p ^{{D\over 2}}\over (2\p )^2}\G (-{\e \over 2})\int d\sp 2 x\left( 
\sqrt{-g} R + \h^{ab}\G_{a\n}^{\r}\G _{b\r}^{\n}+\F \Box \g \right)\cr
&+{4i\p ^{{D\over 2}}\over (2\p )^2 }\G (-{\e \over 2})\int d\sp 2 x 
\sqrt{-g} [(\nabla \F )^2 - \ha\Box\F ]
 ,&(4.10)\cr }$$
$$\li {&\bar A\bar A=D\int d\sp 2 x d\sp 2 y\bar \g ^{ab}(x)\bar \g ^{cd}(y)
 \pa_a^x\pa _{d}^{y} G(y-x)\pa_b^x
\pa_c^yG(x-y)\cr
=&- {i\pi^{{D\over 2}}\over (2\p )^2} D\G (1-{\e \over 2})
B(2+{\e \over 2}, 2+{\e \over 2})\int d\sp 2 x\sqrt{-g}\left(R{1\over \Box }R
+{4\over \e(1+{\e \over 2})}  R\right), & (4.11)\cr}$$
$$\bar A\bar B={i\p^{{D\over 2}}\over (2\p )^2}\int d\sp 2 x \sqrt{-g} R\F ,
 \eqno(4.12)$$
$$\bar A\bar C =-{i\p^{{D\over 2}}\over (2\p )^2}
\int d\sp 2 x \sqrt{-g} \left( R{1\over \Box }R-4R{1\over \Box}(\nabla \F)^2
+2R\F \right),\eqno(4.13)$$
$$\li {\bar B\bar B&=-2{i\p^{{D\over 2}}\over (2\p )^2}\left( \int
d\sp 2 x\sqrt{-g}(R{1\over \Box }R+4 R-2R\F )\right.\cr
&+\left. \G (-{\e \over 2})\int d\sp 2 x
 (\G^{\n}_{c\m}\G^{\m}_{d\n}\h^{cd}+\F \Box \g +\sqrt{-g} (\nabla \F)^2)
 \right) .&(4.14)\cr }$$

From (3.11), (4.2-14) we get the bare effective action
$$\li {{\bar \G}_1=&S-{N-24\over 96\p }\int d\sp 2 x \sqrt{-g} R{1\over \Box }R 
-{N-24\over 24\p \e}\int d\sp 2 x \sqrt{-g} R - {1\over 2\p }\int d\sp 2 x 
\sqrt{-g}R
{1\over \Box } (\nabla \F )^2 \cr 
&-{\p ^{{\e \over 2}}\over 8\p }[ -8\G (-\e/2)+23]\int d\sp 2 x \sqrt{-g}  
(\nabla \F )^2 
+{5\over 4\p }\int d\sp 2 x \sqrt{-g} R\F  \cr
&-{N\p^{\e \over 2}\over 8\p } \G (-{\e \over 2})\int d\sp 2 x\sqrt{-g}
 ((\nabla \F )^2 - \Box\F ) 
+{\p^{\e \over 2}\over 4\p } \G (-{\e \over 2})\int d\sp 2 x\sqrt{-g}(\Box\F
-\f^{-{3\over2}}) 
\cr
& -{N\p ^{\e \over 2}\over 8\p }\int d\sp 2 x \sqrt{-g} \left( R\F 
-R{1\over \Box}(\nabla \F )^2 -
R{1\over \Box }(\nabla f_0)^2 \right),&(4.15)\cr }$$ 
in the large-radius approximation, where $S$ is the classical action given by 
(1.5). The divergent
part of the action (4.15) agrees with the result of \cite{eno}, up to
boundary terms. After making a modified minimal subtraction of the poles in 
(4.15) we get the renormalized one-loop effective action   
$$\li{\G_1 =&S-{N-24\over 96\p }\int d\sp 2 x \sqrt{-g} R{1\over \Box }R
-{1\over \p }
\int d\sp 2 x \sqrt{-g} \left( {1\over 2}R{1\over \Box  }
 (\nabla \F )^2 -{5\over 4}R\F +{23\over 8}(\nabla \F )^2\right)
 \cr
&-{N\over 8\p }\left( \int d\sp 2 x \sqrt{-g}(R\F -R{1\over \Box}(\nabla \F )^2 -
R{1\over \Box }(\nabla f)^2)\right) 
+ O(e^{2\F }) \quad,&(4.16)}$$  
where we have denoted $f_0$ as $f$.

The expression (4.16) is our final result. Since the spherically reduced models
are good approximation for describing the quantum effects of massive black
holes ($M \gg M_{Pl}$) and for $r\gg r_{Pl}$, 
then by taking the large-$N$ limit of (4.16) and neglecting the $O(r\sp{-2})$
terms, we obtain
$$\G\sp{\prime}_1 =S-{N\over 96\p }\int d\sp 2 x \sqrt{-g} \left( R{1\over \Box }R
-12 R{1\over \Box}(\nabla \F )^2  + 12 R\F - 12
R{1\over \Box }(\nabla f)^2 \right) \quad.\eqno(4.17)$$ 
Note that the matter loops contribution is given by the first three terms in
(4.17), while the last term comes from the graviton loops which are induced 
by a non-zero coupling between the matter and the dilaton.
The result (4.17) can be rewritten in the black hole metric
(1.2) as
$$\li{\G\sp{\prime}_1 =S-{N\over 96\p }\int d\sp 2 x &\sqrt{-g} \left( R{1\over \Box }R
-12 R{1\over \Box}(\nabla \F )^2  + 13 R\F + \Box\F {1\over \Box}R - 12
R{1\over \Box }(\nabla f)\sp 2 \right.\cr 
&\left.- (\nabla \F)\sp 2 + 12 \F\Box\F
-12 \Box\F {1\over \Box}(\nabla f)^2 
-12 \Box\F {1\over \Box}(\nabla \F )^2 \right) .&(4.18)\cr}$$
After performing partial integrations one obtains a simpler form
$$\li{\G\sp{\prime}_1 =S-{N\over 96\p }\int d\sp 2 x \sqrt{-g} &\left( R{1\over \Box }R
-12 R{1\over \Box}(\nabla \F )^2  + 14 R\F - 12
R{1\over \Box }(\nabla f)\sp 2 \right.\cr 
&\left.- 13(\nabla \F)\sp 2 
-12 \F (\nabla f)^2 -12 \F (\nabla \F )^2\right) \quad,&(4.19)\cr}$$
but given that in the case of the collapse geometry there is a non-trivial
boundary, these two forms will differ by boundary terms.

\sect{5. Conclusions}

Note that recently two papers have appeared \cite{bh,no}, where
the conformal factor dependent part of the effective action for a spherical
scalar has been computed. This part of the action gives the
trace anomaly, and their results can be written in our normalization as
$$W = - {N\over 96\p }\int d\sp 2 x \sqrt{-g} \left( R{1\over \Box }R
-12 R{1\over \Box}(\nabla \F )^2  + c_3 R\F \right) \quad,\eqno(5.1)$$
where $c_{3}= -4$ in \cite{bh}, while $c_3 = 12$ in \cite{no}.
Our result for $W$ (the first line of (4.19)) differs from (5.1) by 
the presence of $R(1/\Box)(\nabla f)\sp 2$ term. This can be explained by
the fact that 
in \cite{bh} only the matter loops have been taken into account, while in 
\cite{no} the graviton loops have not been taken into account.

Our analysis implies that 
the trace anomaly part of the one-loop effective action is given by 
$$W = - {N\over 96\p }\int d\sp 2 x \sqrt{-g} \left( c_1 R{1\over \Box }R
+c_2 R{1\over \Box}(\nabla \F )^2  + c_3 R\F +
c_4 R{1\over\Box}(\nabla f)\sp 2 \right) \quad,\eqno(5.2)$$
where from (4.19) it follows that $c_1 = 1$, $c_2 = -12$, $c_3 = 14$ and
$c_4 = -12$. Note that the value of $c_3$ is ambiguous, because the term
$R\F$ is equivalent to $\Box\F {1\over \Box} R$ up to boundary terms. 
Therefore the trace anomaly $T$ can be found from
$$2{dW[kg_{\m\n}]\over dk}{\big |}_{k=1} = \int d\sp 2 x \sqrt{-g} T \quad,$$
so that
$$ T = {N\over 24\p } \left( R
-6(\nabla \F )^2  + \ha c_3 \Box\F 
-6 (\nabla f)\sp 2 \right)\quad. \eqno(5.3)$$
   
Note that if one sets $N=1$ in
(4.16) instead of taking the large-N limit, the corresponding
trace-anomaly part will be again of the form (5.2), but now the $c_i$ 
coefficients will be different from the values obtained from (5.3) for $N=1$. 
In particular, $c_1$ will be negative
($c_1 = 1-24 = -23$), due to the ghost contribution $c_g = -26$, which is the
well-known 2d conformal anomaly. This is a generic situation for all relevant
2d dilaton models \cite{mr4}. The resolution of this paradox has been suggested
in \cite{mr4}, where it was pointed out that a resummation of the 
diagrams is a possible way of obtaining the same result as in the large-N
limit. This was based on the fact that in the CGHS case
one can calculate the exact 
one-loop effective action by using the reduced phase space
quantization \cite{m96}. One then obtains the BPP action \cite{bpp},
which differs from the RST action \cite{rst} by a 
$(\nabla \F)\sp 2$ term, and $c_1 = 1$, $c_2 = 0$ and $c_3 = 4$ in both cases. 
The exact action is the same as the large-N limit action, which 
coincides with the matter loops contribution. From the covariant perturbation
theory point of view, this can be explained by the fact that the ghosts serve 
to cancel the loops containing the pure gauge degrees of freedom, which in the 
2d dilaton gravity case are the graviton and the dilaton. Hence it should be 
possible to perform a resummation of the gauge-field and the ghost loop 
diagrams such that one obtains the same values for the $c_{1,2,3}$ coefficients
as those from the large-N action (in the null-dust case $c_1 = 1$, $c_2 = 0$ 
and $c_3 = 2$ \cite{mr4}). We expect that essentially the
same mechanism should work for the spherical scalar case. However,
due to a non-zero dilaton-matter coupling the large-N limit 
is not the same as the matter loops contribution, and therefore
$c_4$ is non-zero. Also note that for conformally invariant theories the 
coefficient of the pole in the divergent part of the one-loop action is the
same as the finite part. In our case this property does not hold any more,
since the classical theory is not conformally invariant, so that a non-zero
value for $c_4$ is not prevented by a symmetry argument.

The value of the coefficient $c_3$ is apparently regularization-scheme
dependent, although there is a further ambiguity in $c_3$ due to appearence
of the term $\Box \F \Box\sp{-1} R$, which is the same as the $R\F$ term, 
up to
boundary terms. Another potential source of ambiguity is the fact that we 
quantize the theory via $(g,\f)$ variables, rather then via the original 
$(\tilde g ,\F)$ variables. In principle the two quantum theories may differ,
so that one can think that a non-zero value for $c_4$ may be related to this
fact.  One can show that within the one-loop background field
formalism, an invertible field redefinition changes the effective action
by the logarithm of the corresponding Jacobian. Calculating this Jacobian
is difficult in general, because one must know the path-integral measure, which
is not known in the general case. However, the calculations in the 
background-field formalism are done with trivial measures, so that a local 
field transformation induces the following Jacobian in our case
$$|J| = \exp\left(\a \d (0)\int d\sp 2 x \sqrt{-g} \F \right)\quad,\eqno(5.4)$$
where $\a$ is a constant. This expression
is equal to one within the dimensional regularization, since $\d (0) =0$.
More generally, one can expect that
$$|J| = \exp\left(\int d\sp 2 x \hat O \F \right)\quad,\eqno(5.5)$$
where $\hat O$ is an operator made from the metric and 
spacetime derivatives. The form of $\hat O$ is constrained by
the dimensionality of the effective action and the diffeomorphism invariance, 
so that
$$\hat O = \sqrt{-g} ( \b R + \g \Box \sp{-1} R \Box + \cdots )
\quad,\eqno(5.6) $$
where $\b$ and $\g$ are constants and
$\cdots$ stand for higher-derivative terms. However, irrespective of the
exact form of $\hat O$, (5.5) can only affect the value of $c_3$. 
Therefore we expect that the value of $c_4$ stays the
same. The best way to check this is to perform the corresponding
calculation with $(\tilde g ,\F)$ variables.    

Given the complete one-loop effective 
action 
we have derived one can start investigating the solutions of the corresponding
equations of motion in order to find the back-reaction effect. An easier task
would be to study the static vacuum solutions along the lines developed in 
\cite{fis}, where the effective action had only the Polyakov-Liouville term.


\end{document}